\documentclass[journal]{IEEEtran}
\IEEEoverridecommandlockouts
\usepackage{cite}
\usepackage{amsmath,amssymb,amsfonts}
\usepackage{algorithmic}
\usepackage[ruled,linesnumbered]{algorithm2e}
\usepackage{scalerel}
\usepackage{multirow} % multirow for table

\usepackage{titlesec}
\usepackage{graphicx}
\usepackage{textcomp}
\usepackage{xcolor}
	
\usepackage{soul} % highlight
\usepackage{amsbsy}%mathrsfs°üÀïÃæµÄ×ÖÌåÔõÃ´¼Ó´ÖµÄÎÊÌâ
\usepackage{gensymb}
\usepackage{enumitem}
\usepackage{amsthm}
\usepackage{amsmath,bm}
\usepackage{caption}
\usepackage{subcaption}

\theoremstyle{remark}

\def\BibTeX{{\rm B\kern-.05em{\sc i\kern-.025em b}\kern-.08em
    T\kern-.1667em\lower.7ex\hbox{E}\kern-.125emX}}
\begin{document}

\title{Intelligent Surfaces Aided High-Mobility Communications: Opportunities and Design Issues
}

\author{Zixuan~Huang, Lipeng~Zhu, and Rui~Zhang

\thanks{Zixuan Huang and Lipeng Zhu (corresponding author) are with National University of Singapore;
Rui Zhang is with The Chinese University of Hong Kong, Shenzhen, Shenzhen Research Institute of Big Data, and National University of Singapore.}% <-this % stops a space
}
\maketitle
\begin{abstract}
Intelligent reflecting/refracting surface (IRS) is envisioned as a promising technology to reconfigure wireless propagation environment for enhancing the communication performance, by smartly controlling the signal reflection/refraction with a large number of tunable passive elements. 
In particular, the application of IRS in high-mobility scenarios can convert wireless channels from fast fading to slow fading, thus achieving more reliable communications.
In this paper, we first provide an overview of the new applications and opportunities of IRS in high-mobility communications. 
Next, we present two practical strategies for deploying IRS to aid high-mobility communications, namely, roadside IRS versus vehicle-side IRS, and compare their different channel characteristics, handover requirements, and deployment costs.
Then, the main issues in designing IRS-aided high-mobility communications, including node discovery, mode switching, beam alignment/tracking, handover, and multiuser scheduling are discussed for both IRS deployment strategies.
Moreover, numerical results are presented to demonstrate the potential performance gains of IRSs in vehicular communications. Finally, new research directions are pointed out for future work.
\end{abstract}

\section{Introduction}
Future wireless networks are expected to support reliable communications under ultra-high mobility for various fast-growing applications, such as high-speed vehicles, high-speed railways, and unmanned aerial vehicles (UAVs) \cite{6g}.
However, the proliferating demands for high-mobility communications (in, e.g., passenger infotainment, autonomous driving, and intelligent transportation systems) may not be fully fulfilled by today’s fifth-generation (5G) wireless networks. 
The Doppler effect caused by the high-speed movement of users and the random scattering in the wireless propagation environment results in
multi-path channels exhibiting random and fast fading over time, which substantially degrades the communication throughput and reliability.
To address this issue, various technologies, such as adaptive modulation/coding, diversity and adaptive equalization, power/rate control, and active beamforming, have been studied and developed to compensate for the wireless channel fading or adapt to fast time-varying channel conditions \cite{high-mo-ass}. 
However, these techniques are usually applied at wireless transceivers, thus being insufficient to fully mitigate the inherent wireless channel impairments for meeting the stringent quality-of-service (QoS) requirements of future high-mobility communications.

Recently, intelligent reflecting/refracting surface (IRS) has emerged as a promising solution to reconfigure the radio propagation environment via real-time control of the signal reflection/refraction in a cost-efficient manner \cite{tut1,tut2,imple}.
Specifically, IRS is a software-controlled metasurface composed of a large number of passive reflecting/refracting elements that can individually tune the phase shift and/or amplitude of the incident signal in real time with low power consumption. Interested readers may refer to \cite{imple} for more details about the state-of-the-art hardware implementation of IRS.
Therefore, IRS is capable of directly controlling the signal propagation to reshape wireless channels, such as bypassing obstacles for establishing virtual line-of-sight (LoS) links, improving multi-antenna/multiuser channel rank conditions by introducing more controllable channel paths, nulling/canceling interference, refining channel statistics/distributions, etc. 

% \begin{figure} 
%     \centering
%      \begin{subfigure}{0.45\textwidth}
%          \centering
%          \includegraphics[width=\textwidth]{config_rs.eps}
%          \caption{Roadside IRS}
%      \end{subfigure}
%      \begin{subfigure}{0.45\textwidth}
%          \centering
%          \includegraphics[width=\textwidth]{config_vs.eps}
%          \caption{Vehicle-side IRS}
%      \end{subfigure}
%      \caption{Illustration of two IRS deployment strategies for high-mobility communications.}
%      \label{config}
% \end{figure}

Particularly, in high-mobility scenarios, the key functionality of IRS is to improve channel statistics/distributions for achieving ultra-reliable communications.
Specifically, an IRS can establish a dominant virtual LoS link between the base station (BS) and a high-mobility user via tunable signal reflection/refraction, which can effectively convert multi-path fast fading (e.g., Rayleigh fading) channels into LoS-dominant slow fading (e.g., Rician fading) channels. 
% This thus helps significantly reduce both the time and frequency selectivities of wireless channels and thereby achieve broadband URLLC for high-mobility users.
It is worth noting that the reflection-based IRS can enhance the communication performance only when both the BS and its served users are located on the same side (i.e., the reflection half-space) of the IRS. 
To expand the communication coverage, the refraction-based IRS can be deployed to assist the BS in serving users on the opposite side of the IRS, while an additional penetration loss may incur when the signal passes through the refraction-based IRS. 
Thus, both reflection- and refraction-based IRSs are generally needed for achieving ubiquitous connectivity in future high-mobility communication systems.

\begin{figure} 
         \centering
         \includegraphics[width=0.43\textwidth]{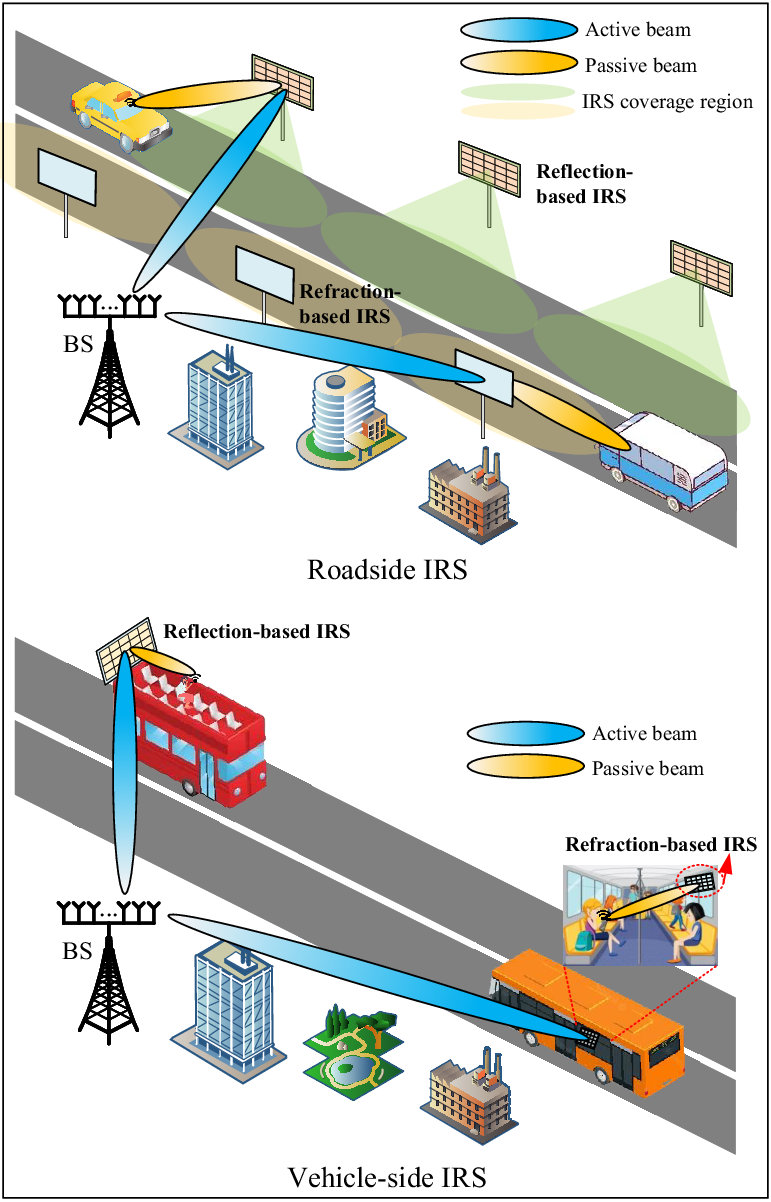}
     \caption{Illustration of two IRS deployment strategies for high-mobility communications.}
     \label{config}
\end{figure}

On the other hand, in the existing literature, IRS is usually deployed at the roadside to enhance the communication performance between the BS and high-mobility users moving on the road\cite{rs_twc,newc1}, as shown in Fig. \ref{config}.
In practice, it is preferred to deploying IRS at designated locations that are close to the high-mobility user or user cluster, which can help reduce the severe product-distance path loss over the IRS-receiving and IRS-reflecting/refracting links between the remote BS and nearby users.
Alternatively, the other practically viable IRS deployment to mitigate the product-distance path loss is the vehicle-side IRS \cite{vs_twc}, as shown in Fig. \ref{config}. 
The vehicle-side IRS can operate either in the reflection mode by placing it inside a vehicle, or in the refraction mode by coating it on the surface of a vehicle.

Although there are promising applications and new opportunities for deploying reflection/refraction-based roadside/vehicle-side IRSs to enhance high-mobility communications, a comprehensive comparison between roadside and vehicle-side IRSs is lacking, which thus motivates this paper to overview their respective design issues and propose promising approaches to tackle them in practice. 
Since the reflection- and refraction-based IRSs follow similar channel models and entail similar signal processing methods, we mainly consider the reflection-based IRS in the rest of this paper for ease of presentation.
The rest of this paper is organized as follows:
Section II compares the roadside and vehicle-side IRS deployment strategies in terms of channel characteristic, handover requirement, and deployment cost, and then elaborates the main communication design issues associated with them as well as potential solutions. 
Section III presents numerical results to compare the performance of  roadside and vehicle-side IRS deployment strategies in a high-speed vehicular communication scenario, as compared to the case without IRS.
Finally, in Section IV, new research directions are pointed out to motivate future investigation.

\section{Roadside IRS Versus Vehicle-side IRS: Comparison and Design Issue}
In this section, we first compare the roadside and vehicle-side IRS deployment strategies for high-mobility communications in terms of the following main aspects.
\begin{enumerate}
    \item \textbf{Channel Characteristic}: 
    For roadside IRS, the IRS-BS channel is usually quasi-static due to the fixed locations of both the BS and IRS, while the user-IRS channel can be rapidly time-varying due to the user's high mobility.
    In contrast, for vehicle-side IRS, the IRS-user channel is typically quasi-static because the IRS and user remain relatively static inside the same vehicle, although they may both move at a high speed with the vehicle; however, the IRS-BS channel may change rapidly over time.
    As such, the corresponding IRS-associated channels under the two IRS deployment strategies both vary rapidly over time in high-mobility scenarios, but with different fast and slow time-varying components, which will be exploited for their respective communication designs as discussed in the sequel of this paper.

    \item \textbf{Handover Requirement}:
    When the IRS-associated channel power is dominant over that of the non-IRS-associated (user-BS direct) channel, the IRS can operate in the user-IRS-BS relaying mode, where the conventional user-BS handover may not be needed, and thus only the user-IRS and IRS-BS handovers are required.
    Since roadside IRSs can only serve the users in their vicinities due to the severe product-distance path loss, a high-mobility user may travel through the coverage regions of multiple IRSs at the roadside within a short period of time, which triggers frequent handovers between the user and adjacent IRSs as well as between these IRSs and the serving BS (i.e., both user-IRS and IRS-BS handovers are required). 
    In contrast, a vehicle-side IRS can easily associate with the onboard users for a long time period, thus requiring much less frequent user-IRS handover. 
    Hence, only the IRS-BS handover is needed over time for the vehicle-side IRS, which is similar to the conventional user-BS handover without any vehicle-side IRS. 
    
    \item \textbf{Deployment Cost}:
    The two IRS deployment strategies require different costs for practical implementation. Due to the limited coverage of each individual IRS, a large number of roadside IRSs should be deployed to guarantee seamless coverage for the road.
    Thus, the deployment cost for roadside IRSs needs to scale with the road length and is likely to be borne by the mobile network provider.
    In contrast, for vehicle-side IRSs, each vehicle requires to be equipped with a single or fixed number of IRSs, and thus the resultant cost scales with the number of vehicles, which is likely to be borne by the vehicle manufacturer. It is worth noting that the practical deployment of vehicle-side IRS may be subject to the size constraint of the vehicle/train, which is more suitable for higher-frequency communication systems with a smaller wavelength and thus IRS element size.

\end{enumerate} 

\begin{table*}
\caption{Comparison of Roadside IRS and Vehicle-side IRS}
\resizebox{\textwidth}{!}
{
\begin{tabular}{|l|ll|ll|l|}
\hline
\multicolumn{1}{|c|}{\multirow{2}{*}{\textbf{Deployment strategy}}} & \multicolumn{2}{c|}{\textbf{Channel characteristics}}                                      & \multicolumn{2}{c|}{\textbf{Handover requirement}}                                                        & \multicolumn{1}{c|}{\multirow{2}{*}{\textbf{Deployment cost}}} \\ \cline{2-5}
\multicolumn{1}{|c|}{}                                              & \multicolumn{1}{l|}{\textit{\textbf{IRS-BS channel}}} & \textit{\textbf{User-IRS channel}} & \multicolumn{1}{l|}{\textit{\textbf{IRS-BS handover}}} & \textit{\textbf{User-IRS handover}} & \multicolumn{1}{c|}{}                                          \\ \hline \hline
Roadside IRS                                                        & \multicolumn{1}{l|}{Quasi-static}                     & Time-varying                       & \multicolumn{1}{l|}{Frequent}                      & Frequent                            & Scales with road length                                    \\ \hline
Vehicle-side IRS                                                    & \multicolumn{1}{l|}{Time-varying}                     & Quasi-static                       & \multicolumn{1}{l|}{Less Frequent}         & Much less frequent                & Scales with vehicle number                            \\ \hline
\end{tabular}
}
\end{table*}
The above comparisons are summarized in Table I. 
The distinct channel characteristics of the two IRS deployment strategies bring different opportunities and challenges in their respective communication designs, as elaborated in the next.
\subsection{Node Discovery and Mode Switching}
% In contrast to the conventional high-mobility communication systems (without IRS) where the BS only needs to discover the high-speed user, 
An initial challenge for roadside/vehicle-side IRS-assisted systems is the node discovery, i.e., the IRS needs to discover high-mobility users in roadside IRS systems, while the BS needs to discover high-mobility IRSs in vehicle-side IRS systems.
To this end, a roadside IRS needs to sense high-mobility users in its vicinity promptly, which however is challenging for fully-passive IRS.
A practically viable solution to this issue is via mounting dedicated sensors on IRS or equipping its controller with multiple antennas \cite{tut1}, so that the IRS can proactively detect the high-speed moving users via sensing their uplink signals.  
Then, the IRS controller can report its detected user information to the BS (via a separate wireless link) for switching from the user-BS direct communication mode to the user-IRS-BS relaying mode.

On the other hand, since the vehicle-side IRS and its served user are co-located inside the same vehicle and remain relatively static, the IRS has sufficient time to discover users in its neighborhood and serve them in the relaying mode.
As such, the BS just needs to discover the high-speed moving IRS instead of the users inside the same vehicle. This can be achieved via either the signal received from the IRS controller or that from the users over the cascaded user-IRS-BS channels in the uplink.

\begin{figure}
\centering
\includegraphics[width=0.43\textwidth]{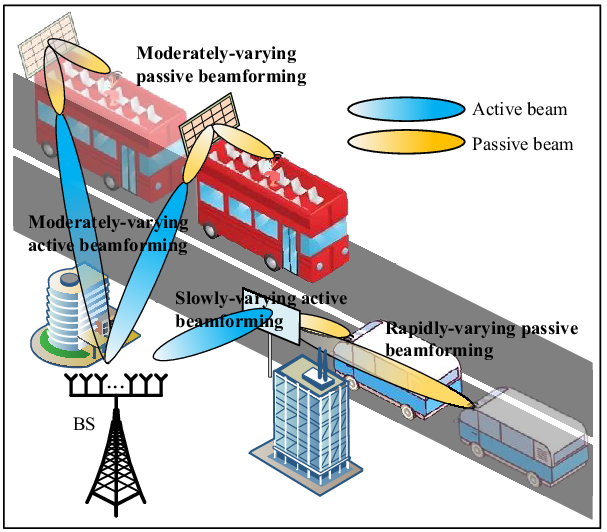}

\caption{Main issues in joint active and passive beam alignment/tracking for roadside IRS and vehicle-side IRS.}
\label{tracking_issues}
\end{figure}

\subsection{Beam Alignment/Tracking}
For both roadside and vehicle-side IRSs, accurate channel state information (CSI) acquisition is essential to achieve the highest IRS passive beamforming gain, which, however, is practically challenging to obtain. This is mainly due to the high-dimensional channel matrices involving a large number of IRS elements (which is generally proportional to the channel estimation time required) as well as the short channel coherence time caused by high mobility.
% Note that it has been shown that if the channel training overhead for IRS-aided systems is too high, the resulting time loss for data transmission may even overwhelm the IRS passive beamforming gain and result in lower achievable rate than that of conventional systems without IRS\cite{rs_twc,vs_twc}, while this issue becomes even more severe for high-mobility communications. 
Nevertheless, it is worth noting that the LoS(-dominant) channel is highly likely to be achieved for user-IRS-BS links by deploying IRS properly in roadside/vehicle-side IRS systems. 
Nevertheless, it is worth noting that for vehicle-side IRS, the LoS(-dominant) channel is very likely to be achieved for the user-IRS-BS link by placing the IRS on the top of a high-speed vehicle/train.
For roadside IRS, the LoS availability of the cascaded BS-IRS-user link is typically higher than that of the vehicle-side IRS, due to the more flexible IRS deployment and more available IRSs on the roadside.
As such, it is practically appealing to perform beam alignment to efficiently track the fast time-varying cascaded user-IRS-BS channel with low training overhead in both cases, as compared to the case with full CSI estimated. 
Recall that the corresponding channels with roadside IRS and vehicle-side IRS have different channel-variation characteristics, which thus motivate us to devise customized two-timescale beam alignment/tracking strategies for them, as illustrated in Fig. \ref{tracking_issues} and discussed as follows in detail.
\subsubsection{Roadside IRS}
After the roadside IRS has detected the nearby high-speed user and informed the BS, the BS can switch to the IRS relaying mode and point its beam towards the IRS's direction based on their quasi-static CSI.
% , which can be efficiently obtained in an offline manner by engaging the IRS controller. 
Hence, the remaining challenge is how the IRS can track the fast time-varying IRS-user channel by adjusting its passive beamforming over time, as shown in Fig. \ref{tracking_issues}. 
Although the conventional beam training method can be applied to search for the best IRS beam in the spatial/angular domain, it may incur prohibitively high training overhead, due to the narrow beams generated by a large number of IRS elements. 
To avoid frequent beam training for high-mobility users, efficient beam tracking methods are crucial, which can be classified into the following two main categories. 
The first category is based on the Kalman filter which models the time-varying yet temporally correlated IRS channel by the Markov process and updates the channel parameters based on the sequential channel measurements over time\cite{track_kal}. 
In contrast, the second category is based on the estimation of the user's velocity, such as angular speed, to predict the angle of arrival (AoA)/angle of departure (AoD) of the fast time-varying IRS-user channel for beam alignment\cite{track_speed}. 
Another possible approach to reduce the beam training overhead is via codebook design with different beam widths, where a trade-off between the beam-width-dependent beamforming gain and beam training overhead exists.

It is worth noting that the IRS beam alignment/tracking can be determined at the BS based on the received signal over the cascaded user-IRS-BS channel, which, however, requires timely feedback from the BS to the IRS for adjusting IRS's reflection coefficients. 
However, the feedback delay in practice may render the designed IRS reflection outdated and less effective as the user-IRS channel varies rapidly due to the user’s high mobility.
A more promising approach is to perform beam tracking at the IRS based on the uplink signal from the user to the IRS sensors/controller, which dispenses with the feedback from the BS.
% However, the channel estimation  accuracy at the IRS sensors/controller is practically limited due to low-cost sensors or insufficient antennas employed at the IRS.
Moreover, the accuracy of beam tracking with roadside IRS can be improved by exploiting the inter-IRS cooperation. 
Specifically, each IRS can collect useful information about the user (such as its moving direction and speed) from neighboring IRSs to improve its channel prediction accuracy and beam tracking performance with reduced sensing overhead. 

Furthermore, broad or flattened IRS beam design is another practically appealing solution to reduce the feedback/sensing overhead with high-mobility users. 
By exploiting the quasi-static roadside IRS-BS channel, the IRS can design a broad beam that covers its served area at the expense of decreased beamforming gain \cite{broad_beam}, which thus avoids the real-time IRS beam alignment/tracking with the high-mobility user. 
However, note that the IRS still needs to discover the user to inform the BS for mode switching in this case since otherwise, the BS cannot adjust its beam to point towards the IRS or IRS controller.

\subsubsection{Vehicle-side IRS}
Under the vehicle-side IRS deployment, both the BS active beamforming and IRS passive beamforming need to be updated over time due to the time-varying IRS-BS channel despite that the IRS-user channel usually remains quasi-static for a long time.    
% under the vehicle-side IRS deployment, due to the relative movement of the BS and IRS, which thus requires the real-time CSI known at both the BS and IRS for designing their beams. 
By exploiting the quasi-static IRS-user channel, the main design issue is how the BS and IRS can track each other by swiftly adjusting their active and passive beamforming, respectively. 
% the fast time-varying IRS-BS channel by adjusting its passive beamforming in real-time. 
It is worth noting that the variation of the IRS-BS LoS channel in the angular domain (i.e., AoA/AoD) is usually slow over time due to the large IRS-BS distance in practice, which can be exploited to reduce the BS/IRS joint beam tracking complexity.

Broad beam design can also be applied to vehicle-side IRS for covering a signal hot-spot inside the vehicle, which avoids frequent IRS beam adjustment for different users when they both move at a high speed.
However, as the cascaded user-IRS-BS channel varies with the time-varying IRS-BS channel, the broad beam at the IRS also needs to be updated over time to match the time-varying IRS-BS channel.

\subsection{Handover}
\subsubsection{Roadside IRS}
Due to the limited coverage of one single roadside IRS, multiple roadside IRSs need to be deployed to ensure the signal coverage over a long distance on the road traveled by the high-speed user, which necessitates frequent handover of each user with different IRSs as well as that of different IRSs with the serving BS at different time (i.e., coordinated user-IRS and IRS-BS handover is required).
However, such coordinated handover is practically challenging to implement due to the following reasons. 
First, the user-IRS handover requires IRS's quick discovery of the high-mobility user and beam alignment to its channel.
Second, immediately after the user-IRS handover, the BS also needs to adjust its beam towards the direction of the newly associated IRS (although such beam can be designed offline based on the quasi-static IRS-BS channel), which may need dedicated feedback from the IRS controller to the BS.
% Considering the above, the handover for roadside IRSs is much more complicated than conventional user-BS handover in high-mobility communications, and this should call for new and efficient designs, so as to make roadside IRS systems practically feasible.
To tackle this challenge, we suggest both centralized and distributed schemes for roadside IRS handover.   
For the centralized scheme, each IRS controller (or sensors) monitors the received signal and reports any discovered user as well as its AoA and moving speed/direction to a local IRS control hub.
Based on the collected information from all IRS controllers, the hub can make handover decisions in a centralized manner and inform each IRS for handover as well as the BS for beam adjustment with different IRSs.
In contrast, for the distributed scheme, each IRS independently senses the signal from any high-mobility user nearby to estimate the user's AoA and moving speed/direction, thereby making its own decision on serving the user and reporting it to the BS. 
The BS then adjusts its beam based on the information gathered from all IRSs. 
Comparing the above centralized and distributed schemes for roadside IRS handover, the former in general provides more efficient handover decisions and BS beam designs due to centralized information processing and optimization, while it also incurs higher implementation cost and may cause larger delay due to the signal exchanges of the control hub with both the IRSs and BS.   
 
\subsubsection{Vehicle-side IRS}
Since the vehicle-side IRS moves along with high-speed users, only the IRS-BS handover needs to be executed, while the IRS serves as a relay for all users, thus avoiding the handover between the users and different BSs.
However, both the IRS and BS are required to change their beam directions during the IRS-BS handover.
For fully-passive IRS (without any sensing capability), both the serving BS and adjacent BSs need to feed the CSI of their corresponding IRS-BS links back to the IRS controller for switching its beam from the direction of the serving BS to that of an adjacent BS, when the handover is conducted. However, if the IRS possesses sensing capability, it can switch its beam more swiftly based on received signals from different BSs in the downlink, thus avoiding any CSI feedback from them.

\subsection{Multiuser Support}
\subsubsection{Roadside IRS}
In practice, there are usually multiple vehicles/users passing by each roadside IRS, which needs to enhance their communications at the same time. 
This thus renders the various design issues for the single-user case discussed previously more difficult to solve in the general multi-user setting. 
In this case, more sensors need to be equipped at each roadside IRS to improve its spatial resolution for separately discovering multiple users at different locations, and then conducting handover and beamforming to serve them efficiently. 
The main new challenge here is to ensure the design being robust to users' locations and their different moving speeds, as well as the potential interference among the users. 
% In particular, how to design efficient IRS beamforming to serve multiple users is practically challenging.   
A low-complexity scheme is to partition each roadside IRS into multiple sub-surfaces, each consisting of a smaller number of reflecting/refracting elements. 
Hence, each sub-surface can be assigned to beam towards one user only at each time and thus multiple users can be served simultaneously if their interference is small thanks to their different AoAs/AoDs to/from the IRS.
However, the channels of users with similar mobility patterns (e.g., those residing in the same vehicle) are practically highly correlated and thus prone to multiuser interference, which may render the surface partition scheme less effective.
To tackle this issue, orthogonal multiple access (OMA) schemes can be applied to serve the spatially correlated users. 
For example, the co-located/spatially correlated users can be served by the same IRS beam in a time division multiple access (TDMA) manner, which thus requires multi-user scheduling and time allocation to balance their performance. 
Another practical approach with even lower complexity than surface partition is to design a broad beam at each roadside IRS which guarantees the coverage of a given region, regardless of the number of users and their locations, and applied jointly with OMA for multi-user scheduling.  
Note that since the IRS-BS channel is usually static and the coverage region of each roadside IRS is fixed, its broad beam does not need to be frequently changed once designed, which thus greatly simplifies the system implementation.
\subsubsection{Vehicle-side IRS}
Since different IRS-user channels are quasi-static for multiple users and the IRS residing in the same vehicle, their CSI can be acquired in a large timescale, based on which the IRS partition scheme/broad beam design proposed previously can be applied with different multiple access methods. 
As a result, the BS and IRS only need to track the CSI of the IRS-BS channel when the vehicle moves, for which the overhead does not scale with the number of users in the vehicle and our previously proposed designs for the single-user case also apply similarly to the multi-user case. 
From the above discussions, it is evident that the vehicle-side IRS systems are easier to implement as compared to their roadside IRS counterparts, especially when multi-user support is required for each IRS. 
\section{Numerical Results}
Numerical results are presented in this section to compare the channel behaviors in the roadside and vehicle-side IRS systems, as well as their performance under a high-speed vehicular communication setup.
In particular, we consider the downlink communication from a static BS to a high-speed vehicle/user. 
The BS is located at $(-150,200,50)$ meters (m) in a three-dimensional Cartesian coordinate system, and it is equipped with a horizontal 16-element uniform linear array (ULA). 
The high-speed user equipped with a single antenna is initially located at $(43.75,0,0)$~m and then moves along the $x$-direction at a speed of $v = 50$ meters per second (m/s) over a traveling period of 0.5 seconds (s). 
For the roadside IRS deployment, two reflection-based IRSs each with 200 reflecting elements are located at $(50,-2,2)$~m and $(62.5,-2,2)$~m, respectively. 
The user selects the nearest IRS for communication at any time, and thus each of the two IRSs serves the user for 0.25~s. 
For the vehicle-side IRS deployment, one reflection-based IRS with 200 reflecting elements is initially located at $(43.75,-2,2)$~m and then moves along with the user.
Note that the fixed IRS-user distance (i.e., 2.8~m) in the vehicle-side IRS case is equal to the minimum IRS-user distance in the roadside IRS case.
% The maximum Doppler frequency is given by $f_{max} = v f_c / c$, where $f_c = 5.9$~GHz and $c = 3\times10^8$~m/s denote 
The carrier frequency is set as 5.9~GHz, as specified by both the 3GPP and IEEE 802.11g standards for cellular vehicle-to-everything applications.
For both IRS deployment strategies, we assume LoS channels for both the IRS-BS and user-IRS links with the path loss exponent equal to 2.
We assume that the channel of the user-BS direct link follows Rayleigh fading with its spatial/time correlation following the Jake’s spectrum and the path loss exponent equal to 3.
The coherence time is set as one-tenth of the maximum Doppler frequency.
The reference channel power gain at a distance of 1~m is set as $-48$~dB. 
We consider the 2-bit controlled discrete phase shifter with 4 equal-interval phase values between 0 and $2 \pi$ for each IRS element.
To evaluate the performance limit of the two IRS deployments, we assume that the perfect CSI of the cascaded user-IRS-BS link is available at the IRS for designing the passive beamforming that maximizes the power of the IRS-reflected channel. It is worth noting that the performance loss due to the maximization of the IRS-reflected channel gain without considering the non-IRS-reflected channel is practically small when the power of the IRS-reflected channel is much stronger than that of the non-IRS-reflected channel.

\begin{figure}
\centering
\includegraphics[width=0.4\textwidth]{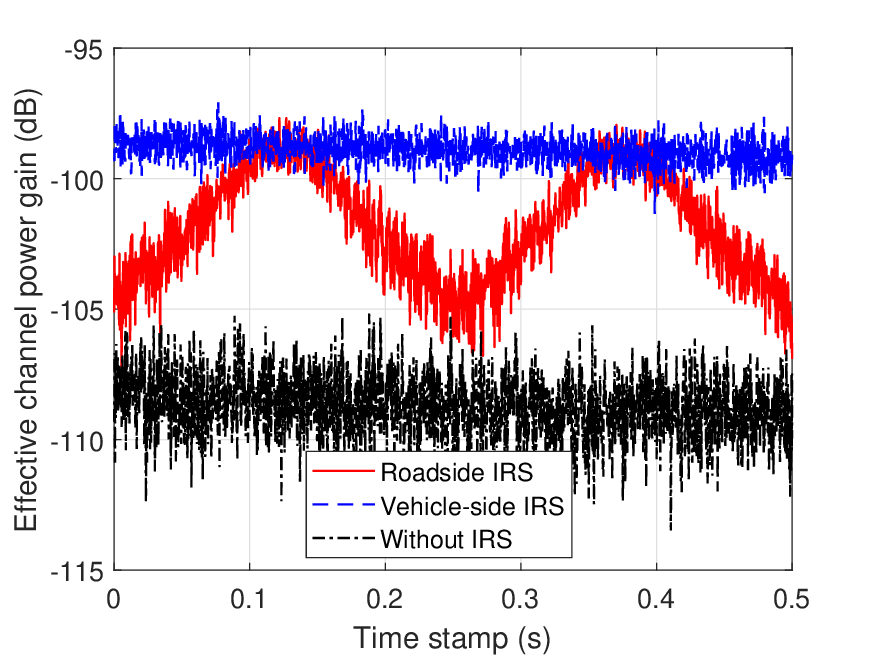}
\caption{A realization of the effective channel gain over time.}
\label{snr-time}
\end{figure}

In Fig. \ref{snr-time}, we show one realization of the effective channel (including both the IRS-reflected and user-BS direct channels) power gains of the roadside IRS deployment, the vehicle-side IRS deployment, and the system without IRS, respectively, over time. 
It is observed that the vehicle-side IRS deployment maintains a much higher channel gain than the other two schemes, and also fluctuates much less over time. 
This demonstrates that the vehicle-side IRS can create a dominant virtual LoS (IRS-reflected) path, thus effectively converting the channel from fast fading in the case without IRS to slow fading.
In contrast, the effective channel gain under the roadside IRS deployment also improves significantly over the case without IRS on average, but it fluctuates drastically over time as the vehicle needs to travel through a coverage ``hole" that lies in the middle of the two adjacent roadside IRSs. 
This result indicates that for roadside IRS systems, it is crucial to design the deployment of adjacent IRSs properly to ensure seamless coverage along the road.

\begin{figure}
\centering
\includegraphics[width=0.4\textwidth]{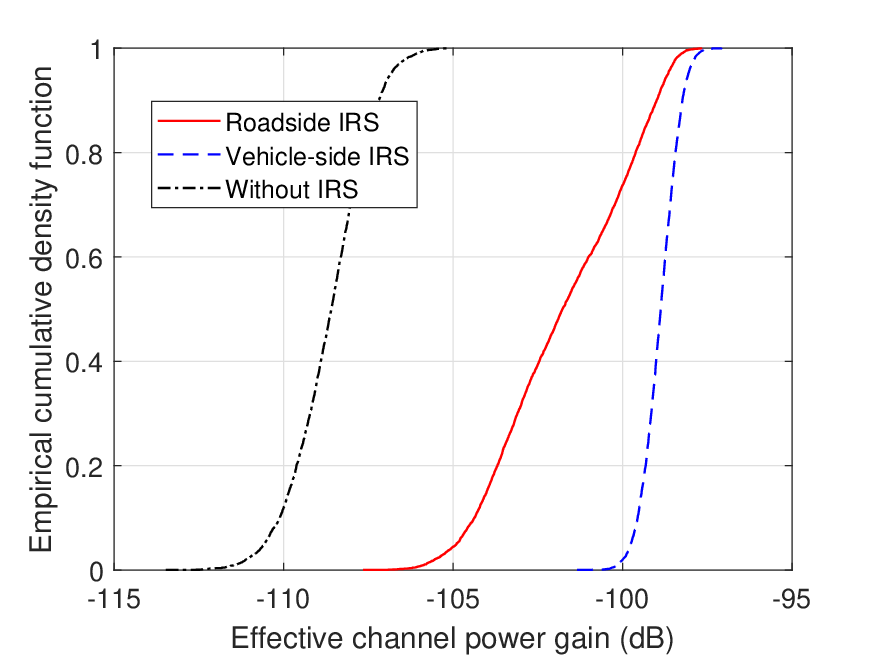}
\caption{Empirical CDF of the effective channel gain.}
\label{cdf}
\end{figure}

In Fig. \ref{cdf}, we compare the empirical cumulative distribution function (CDF) of the effective channel power gains of the roadside IRS deployment, the vehicle-side IRS deployment, and the system without IRS. 
It is observed that the effective channel gain distribution is significantly improved under both IRS deployment strategies as compared to the case without IRS, owing to the high passive beamforming gain provided by IRS. 
Moreover, the vehicle-side IRS deployment not only achieves a higher average channel power gain but also has a smaller channel gain variation as compared to the roadside IRS deployment, because of the constant IRS-user distance and the static channel between them.

\section{Extensions and Future Directions}
% In addition to the aforementioned design issues in IRS-aided high-mobility communications, there are other extensions and corresponding new problems that remain open to be solved. 
% In this section, we highlight some of them to motivate future work.  
% \begin{figure}[ht]
%     \centering
%      \begin{subfigure}[t]{0.4\textwidth}
%          \centering
%          \includegraphics[width=\textwidth]{future_double.eps}
%          \caption{Hybrid IRS Deployment}
%      \end{subfigure}
%      \hfill
%      \begin{subfigure}[t]{0.4\textwidth}
%          \centering
%          \includegraphics[width=\textwidth]{future_activeeps.eps}
%          \caption{Active IRS}
%      \end{subfigure}
%      \hfill
%      \begin{subfigure}[t]{0.4\textwidth}
%          \centering
%          \includegraphics[width=\textwidth]{future_sensing.eps}
%          \caption{IRS-aided Vehicle Sensing}
%      \end{subfigure}
%      \caption{Illustration of other IRS-aide high-mobility systems.}
%      \label{future_fig}
% \end{figure}
% \begin{figure}[ht]
%          \centering
%          \includegraphics[width=0.4\textwidth]{future_work_v.eps}
%      \caption{Illustration of other IRS-aide high-mobility systems.}
%      \label{future_fig}
% \end{figure}
\begin{figure*}[ht]
         \centering
         \includegraphics[width=0.98\textwidth]{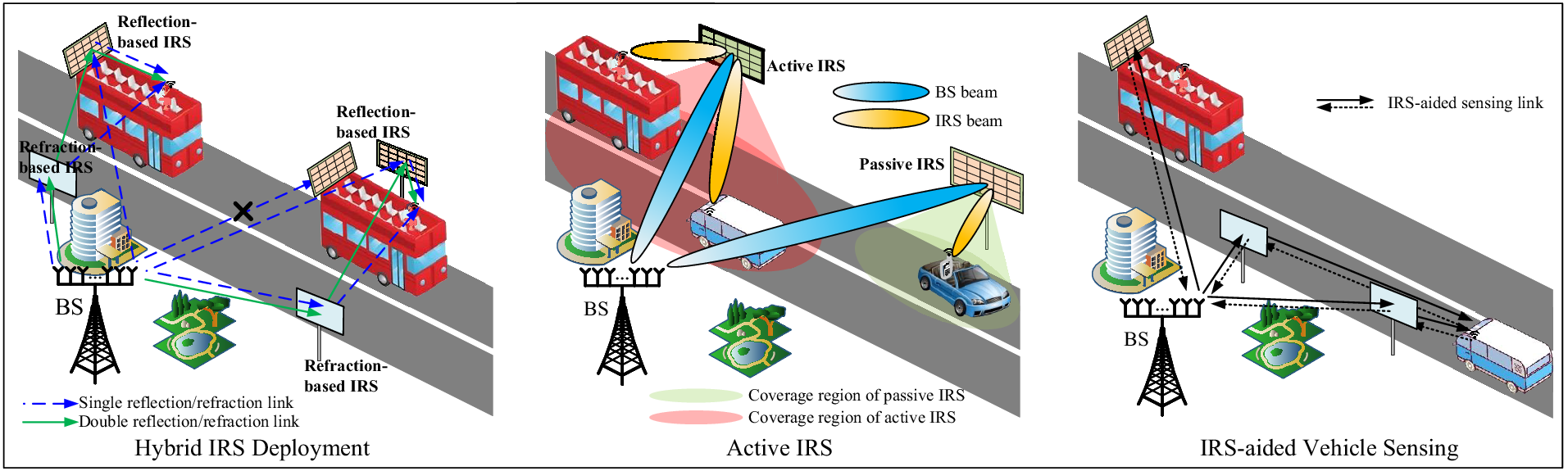}
     \caption{Illustration of other IRS-aided high-mobility systems.}
     \label{future_fig}
\end{figure*}

\subsection{Hybrid Roadside/Vehicle-side IRS Deployment}
In practice, both roadside and vehicle-side IRSs, reflection-based or refraction-based, can be deployed flexibly to improve the communication performance, which is thus called hybrid roadside and vehicle-side IRS deployment. 
In this case, in addition to the IRS single-reflection/refraction user-IRS-BS links, double-reflection/refraction links can be exploited to further enhance the received signal power\cite{double_BX}, as shown in Fig. \ref{future_fig}.
However, to reap the performance gain from the double-reflection/refraction links, efficient CSI acquisition and IRS beam alignment/tracking schemes need to be designed, which is challenging for the high-mobility communication scenarios.  
Moreover, under the hybrid deployment, the roadside and vehicle-side IRSs can complement each other, thus greatly improving the reliability and robustness of high-mobility communications. 
% For instance, when the vehicle-side IRS link is blocked, the roadside IRS can still help maintain an LoS link with the user and BS, as shown in Figure \ref{future_fig}(a), thus providing a higher diversity by IRS selection in practical systems. 
\subsection{Active IRS}
The coverage of a single passive IRS is practically limited due to the severe product-distance path loss in the user-IRS-BS link. 
Alternatively, a new type of IRS, namely active IRS  \cite{cs_active,zy_active}, can be employed to achieve broader coverage and thus reduce the number of IRSs deployed, thanks to its low-cost negative resistance components that are able to amplify the reflected/refracted signal while adjusting its phase, as shown in Fig. \ref{future_fig}. 
To exploit both the reflection/refraction beamforming and power amplification gains offered by the active IRS, the CSI of the cascaded transmitter-IRS-receiver link and additional statistical information on the IRS amplified noise at the receiver are required, which are challenging to acquire due to the typically short channel coherence time in high-mobility communication systems.
% Moreover, to achieve satisfactory communication performance, the IRS active beamforming should be designed carefully to balance the amplifications of signal and noise. 
\subsection{IRS-Aided Vehicle Sensing}
% The rapidly time-varying channels between the BS and high-mobility vehicle-side IRS as well as between the roadside IRS and high-mobility users render IRS-aided high-mobility communication challenging to implement.
The roadside/vehicle-side IRS-aided BS sensing can be invoked to help detect the high-speed vehicle at the BS and thereby resolve some of the challenging issues in IRS-aided communications, such as BS-IRS beam alignment for vehicle-side IRS and IRS-BS handover for roadside IRS, as shown in Fig. \ref{future_fig}.
Specifically, reflection-based vehicle-side IRS can change its reflection direction periodically over time to help the BS sense the direction of the moving vehicle, which also helps the BS and IRS find their jointly optimal beam alignment for communicating with the users inside the vehicle. 
Moreover, the refraction-based roadside IRS can help the BS find the direction of the vehicle by varying its refraction direction periodically, which also helps the IRS hand over to the BS for serving the users inside the vehicle.

\section{Conclusions}
In this paper, we provide a comprehensive overview of the applications of IRS in high-mobility communications. 
We show that IRS is able to overcome the performance limitations of existing high-mobility communication systems, because the virtual LoS path created by the IRS can effectively convert wireless channels from fast-fading to slow-fading and improve significantly their average channel power. 
In addition, we present two practical IRS deployment strategies, namely, roadside IRS and vehicle-side IRS, and compare their communication performance as well as respective design issues, including node discovery, beam alignment/tracking, handover, and multiuser support.
Finally, we point out open problems and new directions for IRS-aided high-mobility communications to inspire future research.

\section*{Acknowledgement}
This work is supported in part by Ministry of Education, Singapore under Award T2EP50120-0024, Advanced Research and Technology Innovation Centre (ARTIC) of National University of Singapore under Research Grant R-261-518-005-720, and The Guangdong Provincial Key Laboratory of Big Data Computing.

% \begin{Acknowledgement}
%     This work is supported in part by Ministry of Education, Singapore under Award T2EP50120-0024, Advanced Research and Technology Innovation Centre (ARTIC) of National University of Singapore under Research Grant R-261-518-005-720, and The Guangdong Provincial Key Laboratory of Big Data Computing.
% \end{Acknowledgement}

% \newpage

\section*{Biographies}

\vskip -3\baselineskip plus -1fil
\begin{IEEEbiographynophoto}{Zixuan Huang} (huang.zixuan@u.nus.edu) is a Ph.D. candidate with National University of Singapore, Singapore.
\end{IEEEbiographynophoto}
\vskip -2.5\baselineskip plus -1fil
\begin{IEEEbiographynophoto}{Lipeng Zhu} (zhulp@nus.edu.sg) is a Research Fellow with the Department of Electrical and Computer Engineering, National University of Singapore.
\end{IEEEbiographynophoto}
\vskip -2.5\baselineskip plus -1fil
\begin{IEEEbiographynophoto}{Rui Zhang} [F'17] (elezhang@nus.edu.sg) is a professor with the School of Science and Engineering, The Chinese University of Hong Kong, Shenzhen, and Shenzhen Research Institute of Big Data. He is also with the ECE Department of National University of Singapore.
\end{IEEEbiographynophoto}

\vfill

\end{document}